\documentstyle[epsfig]{mn}{}

\begin{document}
\def\mpc{h^{-1} {\rm{Mpc}}}
\def\kpc{h^{-1} {\rm{kpc}}}
\def\up{h^{-3} {\rm{Mpc^3}}}
\def\uk{h {\rm{Mpc^{-1}}}}
\def\lsim{\mathrel{\hbox{\rlap{\hbox{\lower4pt\hbox{$\sim$}}}\hbox{$<$}}}}
\def\gsim{\mathrel{\hbox{\rlap{\hbox{\lower4pt\hbox{$\sim$}}}\hbox{$>$}}}}
\def\kms {\rm{km~s^{-1}}}
\def\masa{h^{-1}\rm{M_{\odot}}}
\def\apj {ApJ}
\def\aj {AJ}
\def\aa {A \& A}
\def\mnras {MNRAS}
\title{Environmental Influences on the Morphology \& Dynamics of 
Group Size Haloes}
\author[Ragone-Figueroa \& Plionis]
{
  \parbox[t]{\textwidth}
  {
    Cinthia Ragone-Figueroa${^{1,2}}$ \&
    Manolis Plionis${^{3,4}}$
  }
  \vspace*{6pt}\\ 
  \parbox[t]{15 cm}
  {
    $1$ Grupo de Investigaciones en Astronom\'{\i}a Te\'orica y Experimental, 
    IATE, Observatorio Astron\'omico, Laprida 854, 5000, 
    C\'ordoba, Argentina.\\ 
    $2$ Consejo de Investigaciones Cient\'{\i}ficas y T\'ecnicas de la
    Rep\'ublica Argentina.\\
    $3$ Institute of Astronomy \& Astrophysics, National Observatory of Athens,
    Palaia Penteli 152 36, Athens, Greece.\\
    $4$ Instituto Nacional de Astrof\'{\i}sica Optica y Electr\'onica, AP 51
    y 216, 72000, Puebla, M\'exico.\\
  }
}
\date{\today}

\maketitle

\title{Morphology \& Dynamics of Group Size Haloes} 

\begin{abstract}
We use group size haloes, with masses in the range $10^{13}< M<
2\times10^{14}\masa$, identified with a ``friends of friends'' (FOF) 
algorithm in a concordance $\Lambda \rm{CDM}$ GADGET2 (dark matter only) 
simulation to investigate the
dependence of halo properties on the environment at $z=0$. 
The study is carried out using samples 
of haloes at different distances from their nearest massive {\em cluster} halo, 
considered as such if its mass is larger than the upper
limit of the above halo mass range (ie., $M\ge 2 \times 10^{14} \masa$). 
We find that the fraction of haloes with substructure typically
increases in high density regions. 
The halo mean axial ratio $\langle c/a\rangle$ also increases in
overdense regions, a fact which is true for the whole range of halo
mass studied. This can be explained as a
reflection of an earlier halo formation time in high-density regions, 
which gives haloes more time to evolve and become more spherical. 
Moreover, this interpretation is 
supported by the fact that, at a given halo-cluster distance, haloes 
with substructure are more elongated than their equal mass 
counterparts with no substructure, reflecting that the
virialization (and thus sphericalization) 
process is interrupted by merger events.  
The velocity dispersion of low mass haloes with strong
substructure shows a significant increase near massive clusters with respect 
to equal mass haloes with low-levels of substructure or with haloes
found in low-density environments.
The alignment signal between the shape and the
velocity ellipsoid principal axes decreases going from lower to 
higher density regions, while such an alignment is stronger for haloes
without substructure.
We also find, in agreement with other studies, a tendency of 
halo major axes to be aligned and of minor axes to lie roughly perpendicular 
with the orientation of the filament within which the halo is 
embedded, an effect which is stronger in the proximity of the massive clusters.
\end{abstract}


\section{INTRODUCTION}
\label{introduccion}
According to the current Cold Dark Matter ($\rm{CDM}$) paradigm, 
haloes emerging from
a Gaussian primordial density fluctuation field, assemble through gravitational
processes to form larger systems which eventually virialize. 
These structures evolve in a hierarchical fashion aggregating smaller
mass systems, flowing out of voids and along filaments, 
giving rise to deep potential wells, the cluster of galaxies. 
The role that the environment plays in modifying the properties of the
smaller systems, such as galaxies, is being exhaustively studied and
it is well known that many of the observed galaxy properties correlate
strongly with environment (eg. Dressler 1980; Goto 2003; see 
Boselli \& Gavazzi 2006 for a recent review).
The properties of the galaxy group-size haloes within which they are embedded
could also vary as a function of environment, since mergers and
tidal interactions are more probable in high density environments.

The proximity of a galaxy or a group-size halo to a massive
attractor, like a cluster, and the corresponding strong gravitational 
interactions not only with the cluster itself but also with its local 
surrounding, which is denser near the cluster,
might affect halo properties such as, among others, shape, size, 
concentration, orientation, 
velocity dispersion, amount of substructure and internal alignments
(for examples of such observational evidences, see Schuecker et al. 2001,
Plionis \& Basilakos 2002, Plionis 2004, Martinez \& Muriel 2006).
%which in turn could influence the properties of galaxy members.
Quantifying such effects in numerical simulations and understanding
their significance could help understand the physical processes that
act to determine the properties of galaxies as a function of their environment.

However, one should also remember that
differences between galaxy and group haloes, in high and low 
density regions, could also arise as a natural consequence of cosmological 
initial conditions, like halo formation time (eg., Gottl\"ober, Klypin
\& Kratsov 2001; Sheth \& Tormen 2004) or halo  spin generation
efficiency as a function of local density (Lee 2006).

Various recent studies have applied environment detecting algorithms in an
attempt to characterize the diversity of cosmic environments from voids, to
walls, filaments and clusters and thus facilitate the study of
environmental effects on galaxy, group and cluster properties.
 Such algorithms are based on a variety
of pattern recognition techniques from
the simplest local overdensity measures to more elaborate techniques
based on second-order local variations of the density field 
(eg. Pimbblet 2005; Colberg et al. 2005; Stoica et al. 2005; 
Arag\'on-Calvo et al. 2006; Hahn et al. 2006)

Lemson \& Kauffman (1999), explored the effect of environment on 
different halo properties like their mass function, 
concentration parameter, formation redshift, spin parameter and shape 
and found that halo mass is the only property
that correlates significantly with local environment. 
It is important to note that the variation of the halo mass function 
in different environments, ie., the fact that 
high mass haloes are under-represented and over-represented in low 
and high density regions, respectively, suggests that any apparent 
dependence of halo properties on the environment, could be a
consequence of the dependence of these properties on halo mass.
Therefore, one needs to
disentangle the two dependencies and to perform any environmental
dependence study as a function of halo mass as well.

A large number of recent studies on the environmental
effects on a variety of halo properties, like halo shapes, spin, alignments
velocity dispersion, and for different mass halo
ranges, have been presented (eg. Faltenbacher et al. 2002; 
Einasto et al. 2003; Ragone et al. 2004; Einasto et al. 2005; 
Avila-Reese et al. 2005; Hopkins, Bahcall \& Bode 2005; Basilakos et al. 2006; 
Plionis et al. 2006; Altay et al. 2006; Maulbetsch et al. 2006;
Arag\'on-Calvo et al. 2006; Hahn et al. 2006). However,
results of different studies are not always in agreement with each
other, a fact that could be due to different quantifications of the
environment or due to different analysis tools.
For example, Einasto et al. (2005) found that group and cluster size
DM haloes, in high density regions, have smaller eccentricities 
(are more spherical) than in low density regions, while Kasun \&
Evrard (2005) have found no such dependence. 
Similarly, Avila-Reese et al. (2005) and  Hahn et al
(2006) have found a dependence of galaxy-size DM haloes shapes 
on environment but again no such obvious dependence for larger DM haloes.

An interesting property to study as a function of environment, is the internal 
alignment between the principal axes of the shape and
velocity anisotropy ellipsoids, which can be
considered as an indication of relaxation in a system where
the shape is supported by internal velocities (eg. Tormen 1997). 
Kasun \& Evrard (2005)
and Allgood et al. (2006) found for cluster-size 
DM haloes a good such alignment, although no investigation in 
different environments has been reported.
Furthermore, the external alignment between
DM halo axes or angular momentum and the orientation of the filament 
in which the 
halo is embedded is of interest. Bailin \& Steinmetz (2005) found a very 
strong tendency for the halo minor axis to lie perpendicular to the large 
scale filament, but a much weaker tendency for the major axis to be oriented 
parallel to it.  They also found that the group and cluster size 
halo angular momenta lie perpendicular
to the large scale filaments while that of galaxy-size haloes tend to lie
parallel to them. This suggests that group-size DM haloes acquire most of their 
angular momenta from mergers along the filament direction.
Avila-Reese et al. (2005), in turn find a decreasing alignment signal between
minor axis and angular momentum of galaxy-size DM haloes going from overdense
to underdense regions.

In this work we attempt to give a new insight in the behavior of
group-size DM halo properties (shape, velocity dispersion, internal
and external alignments) as a function of environment, 
taking special care to disentangle their correlation with halo mass,
as mentioned before. We also divide our sample of group-size haloes
according to the amount of substructure that they have, in order to infer 
if mergers and/or gravitational tidal interactions 
play a significant role in shaping the DM halo morphological and dynamical 
properties.

The outline of this paper is as follows. In section \ref{simulacion} we 
describe the numerical simulation method, the halo identification
procedure and the research methodology that we
will follow.
In section \ref{formaysubs}, we present the methods for the
computation of the shape and velocity tensors, the angular momentum and
alignment measures and finally we present a thorough study for the
quantification that we use to determine the halo substructure. In section 
\ref{environment} we study the dependence of halo shape and dynamics
on environment while in section \ref{alin} 
we present the corresponding study
of internal and external halo alignments.
Finally, we summarize our results and draw our conclusions
in section \ref{conclusiones}. 

\section{NUMERICAL DATA \& RESEARCH METHODOLOGY}
\label{simulacion}
The numerical simulations used in this work were performed using the 
GADGET2 code (Springel 2005)
with dark matter only. This parallel code was ran in a Beowulf cluster with
$32$ Intel Xeon processors (3.06 GHz).
The cosmological parameters used
correspond to a flat cosmological model with a non-vanishing cosmological 
constant (${\rm \Lambda CDM}$): $\Omega_m=0.3$, $\Omega_\Lambda=0.7$, 
$\sigma_8=0.9$, $h=0.72$, where $\Omega_m$ and  $\Omega_\Lambda$ are the 
present day matter and vacuum energy densities in units of the critical density,
$\sigma_8$ is the present linear rms amplitude of mass fluctuation in 
spheres of $8 \mpc$ and $h$ is the Hubble parameter in units of 
$100\rm{km~s^{-1} Mpc^{-1}}$. 
The initial conditions are generated with the GRAFIC2 package 
(Bertschinger 2001), which also computes the transfer function as described 
in Ma \& Bertschinger (1995).

The main simulation was run in a cube of size $L=500 \mpc$, using $512^3$ 
particles. The particle mass is $\sim 7.7\times10^{10}\masa$ and the force 
softening length is $\epsilon = 100 \kpc$.
Individual particle time-steps are chosen to be proportional to the 
square root of the softening length over the acceleration $\mathbf a$:
$\Delta t_i = \sqrt{2 \eta \epsilon/|\mathbf a|}$. We set the dimensionless
parameter which controls the accuracy of the time-step to be $\eta = 0.02$.

The haloes were identified using a FOF algorithm with a linking length
$l=0.17$ times the mean inter-particle separation.
Given the proposes of this work, we only use haloes with at least 130
particles, ie., with masses greater than $10^{13}\masa$. Note that this 
halo finder does not identify (for a given linking length) sub-haloes 
belonging to larger {\em parent} haloes. For the purpose of our study, 
however, we will consider haloes with and without substructure (see further 
below), which in effect correspond to those haloes with or without 
relatively massive sub-haloes.

The resulting sample of $\sim 58000$ haloes was split in two
subsamples: haloes with masses $ M> 2\times 10^{14}\masa $ are
considered as {\em clusters} (in total 1598 haloes), 
whereas haloes in the range 
$10^{13}\masa <M< 2\times10^{14}\masa$ are considered as {\em groups} 
(56699 haloes).

\begin{figure}{}
\epsfxsize=0.5\textwidth
\centerline{\epsffile{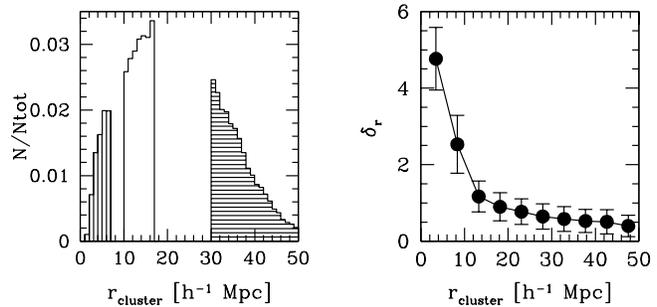}}
\caption{{\em Left panel:}  Fraction of haloes in the H$_{0-7}$ (vertical dashed 
histogram), H$_{10-17}$ (empty histogram)  and
H$_{30-50}$ (horizontal dashed histogram) sample.
{\em Right panel:} The median density contrast $\delta(\rm r)$ 
computed in spheres of radius $8 \mpc$, as a function 
of $r_{\rm cluster}$ distance.  Error bars represent 
the 33\% and 67\% quantiles of the corresponding distribution.}
\label{sobredens}
\end{figure}

In order to investigate the role that environment
plays in determining halo properties,
we find for each halo the distance to its nearest cluster and divide the 
halo sample in three subsamples 
according to this distance $(r_{\rm cluster})$:
\begin{itemize}
\item Small distance subsample: $r_{\rm cluster}<7\mpc$ (H$_{0-7}$ hereafter,
$\sim$ 8\% of the group sample). 
\item Intermediate distance subsample: 
$10\mpc<r_{\rm cluster}<17\mpc$ (H$_{10-17}$ hereafter, $\sim$ 21\% of the 
group sample).
\item Large distance subsample: 
$30\mpc<r_{\rm cluster}<50\mpc$ (H$_{30-50}$ hereafter, $\sim$22\% of the 
group sample).
\end{itemize}
The left panel of Figure \ref{sobredens} shows the fraction of haloes in each 
one of the previously defined subsamples.
In order to ensure that $r_{\rm cluster}$ is defining accurately the 
environment, we also compute for our haloes at different distances from the
clusters the corresponding density contrast 
$\delta(\rm r) = \rho(\rm r) / \bar{\rho} -1$, where
$\rho(r)$ is the density in a sphere of radius $8 \mpc$ around the
halo centers and $\bar{\rho}$ is the mean matter background density.
Results are shown in the right panel of 
Figure \ref{sobredens} from which it is obvious that indeed 
the halo distance to its nearest cluster is related to the
overdensity in which the halo is embedded.

Given the significant effect that mergers and interactions can have 
on the shapes and alignments of haloes, and the fact that
in overdense regions the halo mass function is skewed toward 
the high-mass end (Lemson \& Kauffmann 1999), we will present our 
results as a function of their:

\noindent
(a) halo mass, 

\noindent
(b) environment, determined by $r_{\rm cluster}$, and

\noindent
(c) halo dynamical state, determined by the Dressler \& Shectman
(1998) method (see next section).

Therefore, any dependence
of the halo properties on environment will be disentangled from the 
mass function effects, and will not be attributed to the overabundance of 
high mass haloes in overdense regions.

For the purpose of testing the robustness of our substructure
determination procedure to variations
of the simulation resolution and box size, we
also run (a) one simulation with the same resolution as the main
simulation but in a 8 times smaller box, 
ie., evolving $256^3$ particles in a $L=250 \mpc$ side box (LR
hereafter), and (b) a higher resolution simulation obtained by 
re-simulating the $L=125 \mpc$ 
central box of the former with $256^3$ particles and $\epsilon = 50 \kpc$  
(HR hereafter), reaching a particle mass resolution of $9.7\times10^9\masa$. 
We identify haloes in both the LR and HR simulations using a linking 
length $l=0.17$ times the mean inter-particle separation as in our main
simulation.

\section{DETERMINATION OF HALO SHAPE, ALIGNMENTS \& DYNAMICAL STATE}
\label{formaysubs}

\subsection{Parameter Definition}
The shape of haloes, modeled as ellipsoids, is determined by
diagonalizing their inertia tensor:
\begin{equation}
I_{ij} = \sum_N x_{i,n} x_{j,n} \;,
\end{equation}
where $N$ is the number of particles in the halo and $x_{i,n}$ is the
$i^{th}$ component of the position vector of the $n^{th}$ particle relative 
to the halo center. The principal axes of the fitted ellipsoid
($a,b,c$ with $a\geq b\geq c$) are related to the square root of the 
eigenvalues of the inertia tensor.
The corresponding eigenvectors provide the directions of the
principal axes of the fitted ellipsoid.

Similarly, velocity moments are obtained by diagonalizing the
velocity anisotropy tensor:
\begin{equation}
V_{ij} = \sum_N v_{i,n} v_{j,n} \;,
\end{equation}
where $v_{i,n}$ is the
$i^{th}$ component of the velocity vector of the $n^{th}$ particle relative 
to the halo center of mass velocity. Note that
$a_{\rm vel} \geq b_{\rm vel} \geq c_{\rm vel}$ will denote the 
major, middle and minor axes of the velocity ellipsoids, respectively.

We compute the specific angular momentum of each halo containing N 
particles as:
\begin{equation}
{\mathbf L} = \frac{1}{N}\sum_{N} {\mathbf r_i} \times {\mathbf v_i} ,
\end{equation}
where ${\mathbf r_i}$ and ${\mathbf v_i} $ are the position and
velocity vectors of the particle $i$ relative to the halo center of mass. 

The various alignments between different pairs of vectors,
representing either the principal axes of the halo density and velocity 
ellipsoid, the halo angular momentum or the direction to a neighboring
cluster halo, will be estimated by the
mean of the distribution of $|\cos(\theta)|$, where $\theta$ is the angle 
between the directions of any two vectors, ${\bf \hat{v}_1}$ and ${\bf
  \hat{v}_2}$, we are interested
in. Therefore, 
\begin{equation}
\cos(\theta) = {\bf \hat{v}_1} \cdot {\bf \hat{v}_2} \;\;.
\end{equation}

\noindent
Perfect alignment and anti-alignment 
correspond to $|\cos(\theta)|=1$ and 0, respectively,
whereas for the random three-dimensional case the expected
distribution mean value is $\langle |\cos(\theta)|\rangle=0.5$.

Finally, we use the Dressler \& Shectman (1998) algorithm to
estimate the amount of substructure in haloes. 
Briefly, this method determines the 
mean local velocity $\langle \bf v_{\rm loc} \rangle$ 
and the local velocity dispersion $\sigma_{\rm loc}$ of the nearest $n$ 
neighbors from each halo particle $i$
and compares them with the mean velocity, $\langle {\bf V} \rangle$, 
and the velocity dispersion, $\sigma$, of the whole halo of $N$ particles, 
defining the following measure:
\begin{equation}
\delta_i^2 = \frac{n}{\sigma}[(\langle \bf v_{\rm loc}\rangle- \langle
  \bf V\rangle)^2+(\sigma_{\rm loc}-\sigma)^2] \;,
\end{equation}
where
\begin{equation} 
\sigma^2_{\rm loc} = \frac{\sum_{n}({\bf v}_{\rm loc}- {\bf v}_{\rm i})^2}{n-1} 
\end{equation} and
\begin{equation} 
\sigma^2 = \frac{\sum_{N}(\langle {\bf V}\rangle- {\bf v}_{\rm i})^2}{N-1} \;.
\label{form-sig}
\end{equation}
 
A quantification of the substructure present in a halo is given by the
so-called  $\Delta$-{\em deviation}, which is the sum of the
individual $\delta_i$'s over all halo particles $N$:
\begin{equation}
\Delta = \frac{\sum_{N} \delta_i}{N} \;.
\end{equation}
The larger the $\Delta$-deviation the stronger is the halo
substructure.
This statistic depends on the number of nearest neighbors $n$ which is 
used in the analysis, and as we verified on the number of particles used to
resolve a halo as well.

We have computed the $\Delta$-deviation using two different
values of $n$: (a) $n=25$ as in 
Knebe \& M\"uller (1999) and (b) $n=N^{1/2}$ as in Pinkney
et al (1996) and find similar results.

\subsection{Random and Systematic Parameter Uncertainties}
We investigate the uncertainty introduced by resolution effects
in the determination of our morphological and dynamical halo parameters.
The fact that low mass haloes are resolved by a smaller number of 
particles with respect to higher mass haloes would inevitably create 
a random or possibly even a systematic deviation from 
their nominal values. To investigate these uncertainties 
we use a procedure similar to that of Avila-Reese et al. (2005). 

We perform 100 realizations of each massive halo (having more than 5000 particles) 
which we resolve selecting randomly the same number of particles as that of 
the lowest mass haloes used in our analysis (ie., 130 random
particles). For each realization 
we then compute the halo $c/a$ axis ratio,  velocity dispersion, 
velocity-shape major axes mis-alignment angle,
${\bf \hat{a}} \cdot {\bf \hat{a}}_{\rm vel}$, and minor axis-angular momentum 
mis-alignment angle, ${\bf \hat{c}} \cdot {\bf \hat{L}}$.

\begin{figure}
\epsfxsize=0.5\textwidth
\hspace*{-0.5cm} 
\centerline{\epsffile{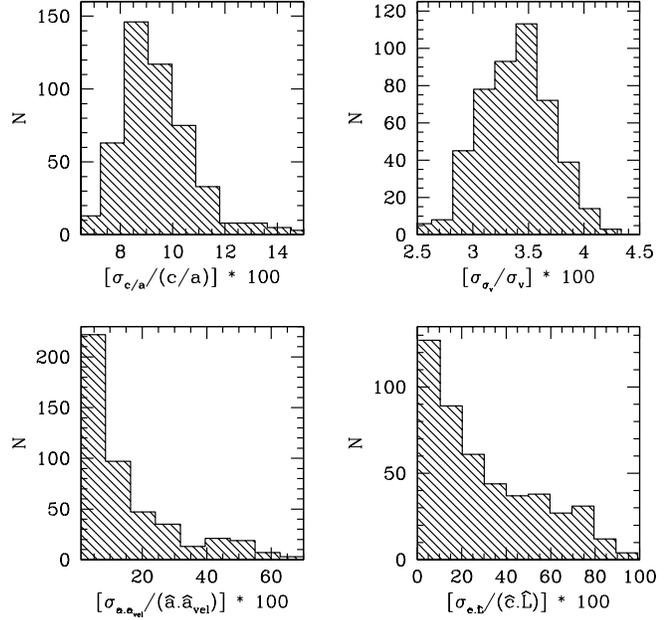}}
\caption{
$1\sigma$ error distribution of the various dynamical and
  morphological parameters of haloes with more than 5000 particles but
sampled with only 130 random particles (100
realizations are used). 
The distributions of the halo axial ratios, velocity dispersions,
velocity-shape major axes mis-alignment angles and minor axis-angular
momentum mis-alignment angles are shown in the 
{\em Top left}, {\em Top right}, {\em Bottom left} 
and {\em Bottom right panels}, respectively.}
\label{errores}
\end{figure}

The distributions of the $1\sigma$ deviations from their nominal value,
plotted in Figure \ref{errores}, have mean values of
$\sim 9\%, 3.5\%, 16\%$ and $30\%$, respectively for the
$c/a$ axis ratio (top left panel), velocity dispersion (top right panel),
velocity-shape major axes mis-alignment angle  (bottom left panel) 
and minor axis-angular momentum mis-alignment angle (bottom right
panel). It is evident that the uncertainties are quite small,
especially of the halo velocity dispersion, except for the
mis-alignment angle, ${\bf \hat{c}} \cdot {\bf \hat{L}}$.  

\begin{figure}
\epsfxsize=0.5\textwidth
\hspace*{-0.5cm} \centerline{\epsffile{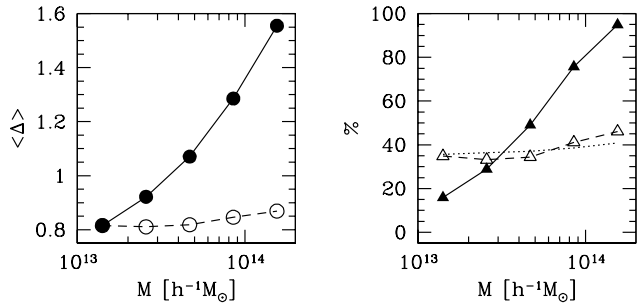}}

\caption{
{\em Left panel:} The dependence of the mean $\Delta$-deviation 
substructure index on halo mass. 
The filled circles, connected with the solid line, represent 
the mean $\Delta$-deviation computed using all the halo member
particles, whereas the corresponding values choosing to have the same
halo resolution, ie.,  $\sim 130$ 
random particles per halo ($\Delta_{\rm ran}$),
is shown as open circles connected with the dashed line. 
The estimated bootstrap uncertainties are smaller than the size of the symbols.
{\em Right panel:} The percentage of haloes with
$\Delta$-deviation higher than the mean value (ie., haloes with
substructure), as a function of halo mass. 
The solid line and filled symbols 
correspond to haloes selected using the global
distribution of $\Delta$-deviations, estimated using all halo
particles, irrespective of the halo mass (variable halo particle resolution). 
The dashed line and open symbols correspond to haloes selected using the global
distribution of $\Delta_{\rm ran}$-deviations, estimated using $\sim
130$ particles per halo, irrespective of the halo mass (same halo particle
resolution). While the dotted line shows the percentages of haloes 
with $\Delta>\langle \Delta_i \rangle$, evaluated within the $i^{\rm th}$ 
bin of halo mass and using all halo particles. 
}
\label{sub}
\end{figure}

It is also possible that resolution effects do not only introduce a random 
error on the nominally defined shape and dynamical halo parameter.
For example, for the case of the substructure index, $\Delta$, 
a pronounced trend is apparent with $\langle \Delta\rangle$ increasing 
with halo mass. In Figure \ref{sub} (left panel) we show results based on the 
$n = N^{1/2}$ case (filled circles \& solid line). 
There is an apparent monotonic increase of $\langle \Delta\rangle$
with halo mass.
Furthermore, in the right panel of Figure \ref{sub} we show 
with the solid line, the percentage of haloes with
$\Delta$-deviation higher than the mean of the $\Delta$ 
distribution of all haloes ($\langle \Delta \rangle= 0.98$). Again, we see 
a monotonic increase of the fraction of haloes having
substructure as a function of mass, with the most massive haloes
appearing all to be substructured.
These results create a suspicion that they could be due to the lower 
resolution with which the low mass haloes are resolved. 
To investigate the resolution issue we perform two tests: 
\begin{itemize}
\item we compare the $\Delta$-deviation index for the matching haloes 
of our HR and LR simulations, and 
\item we recompute $\Delta$ for all mass haloes, but using only $130$ 
randomly selected particles per halo (i.e. the same number resolution 
as in the smaller haloes).
\end{itemize}

\begin{figure}{}
\epsfxsize=0.5\textwidth
\centerline{\epsffile{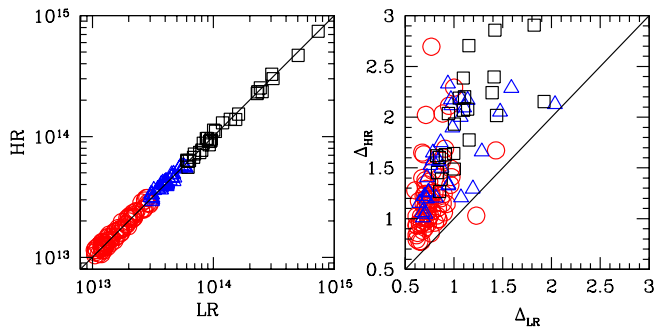}}
\caption{{\em Left panel:} Comparison of the masses of matching 
LR and HR haloes: open circles, triangles and squares correspond to
haloes with masses:
$10^{13}<M< 3\times 10^{13} \masa$,
$3\times 10^{13}<M< 6\times 10^{13} \masa$ and
$6\times 10^{13}<M< 10^{15} \masa$, respectively.
{\em Right panel:} The $\Delta$-deviation comparison of the matched LR
and HR haloes. }
\label{match}
\end{figure}

Regarding the first test we select those pairs of haloes which match,
in position and mass, in both the LR and HR halo samples.
The masses of these matching haloes are allowed to differ only by 5\%,
so as to ensure that we will compare properties of the same haloes with
only difference their resolution
($\sim$ 8 times more particles in the HR matching haloes). 
We divide the matching haloes sample in three subsamples according
to their masses, which are compared in the left panel of Figure \ref{match}.
In the right panel we compare their corresponding $\Delta$-deviation 
values and in all cases we find that $\Delta$ is significantly larger
when computed in the higher resolution haloes, 
a fact which is further enhanced for the more
massive haloes. This result verifies our suspicion that resolution
effects could be the cause of the monotonic increase of $\Delta$ with
halo mass.

We now continue with our second test and derive for each halo a new
$\Delta$-deviation index ($\Delta_{\rm ran}$) computed by using the same
number of particles (130), randomly selected, in each halo independent of
its mass. In this way we impose the same particle resolution on all
haloes. Note that we use
as $\Delta_{\rm ran}$ the average over many realizations of the random
particle selection process.
The results of this procedure show that 
$\Delta_{\rm ran}$ is systematically smaller
than when using all the particles in the haloes, and increases very
weakly with halo mass, as shown by the dashed line in both the left
and right panels of Figure \ref{sub}.
Again these results indicate the importance of resolution effects in
quantifying the amount of halo substructure.

From our previous study we have realized that 
the whole distribution of $\Delta$-deviations shifts to higher
values as a function of resolution. This translates to a shift at
higher values of the $\Delta$-deviation distribution as a function of
halo mass, within the same simulation.

In the right panel of Figure \ref{sub} we also show, as the dotted
 line, the percentage of haloes as a function of halo mass, having
their $\Delta$ estimated using all halo particles,
but then selecting those with $\Delta$
larger than the mean, $\langle \Delta_{\rm bin}\rangle$, 
of the distribution within each bin of halo mass.
It is evident from this plot that this case and the case based
 on $\Delta_{\rm ran}$ give equivalent
results. Had we not taken into account the
resolution effects we would have erroneously concluded that almost 
all massive haloes have strong substructure and that the opposite was
true at the low halo mass end. 

Given the computation of the substructure $\Delta$-deviation 
index is more robust if all halo member particles are
considered, we will refer from now on to haloes having substructure 
as those with 
$\Delta > \langle \Delta_{\rm bin}\rangle$ in the specific mass range bin
which they belong.

The fraction of
haloes with substructure, as defined before, are $\sim 45\% \pm 5\%$ for
the H$_{0-7}$ haloes and 35\%$\pm 3\%$ for the H$_{10-17}$ and 
H$_{30-50}$ haloes, with only a weak dependence on halo mass. 
If, however, we select haloes 
nearer to massive clusters (ie., $0<r_{\rm cluster}<4 \; h^{-1}$ Mpc), the
fraction of low-mass haloes ($M<2\times 10^{13} \;h^{-1} M_{\odot}$)
with substructure grows to 65\% $\pm 3\%$.

Note that haloes that went through a recent merger will have a higher 
$\Delta$-deviation value with respect to those that are either 
isolated or had no recent merger event, thus having more time to
virialize. 

\begin{figure}{}
\epsfxsize=0.5\textwidth
\centerline{\epsffile{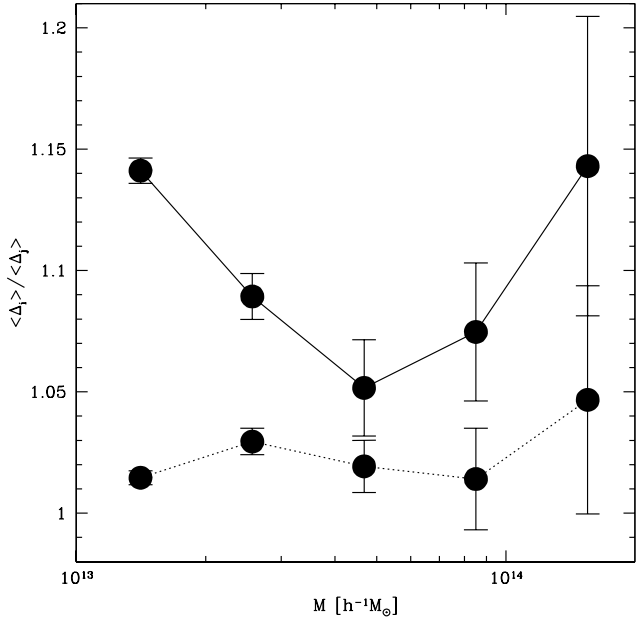}}
\caption{ 
$\Delta$-deviation ratios as a function of halo mass and environment.
The solid line represents the ratio: $\langle \Delta_{\rm H_{0-7}}\rangle/\langle \Delta_{\rm H_{30-50}} \rangle$ 
and the dotted line the ratio:
$\langle \Delta_{\rm H_{10-17}}\rangle/\langle \Delta_{\rm H_{30-50}} \rangle$.
Error bars are based on the propagation of the individual
$\Delta$-deviation uncertainties.}
\label{sub_dist}
\end{figure}

In Figure \ref{sub_dist} we present the dependence of the 
substructure index on the environment and halo mass. 
Solid and dotted lines stand for the ratios of $\Delta$ in the H$_{0-7}$ 
and H$_{10-17}$ samples, normalized to the most distant sample (H$_{30-50}$).
%which cancels the mentioned bias on the $\Delta$ index.
As expected, we find haloes in the vicinity of massive clusters
(solid line) to have larger $\Delta$ values (for their mass range)
with respect to distant haloes (dotted line),
presumably due to the higher merging rate and due to the 
stronger tidal field, found around overdense regions. 

\section{ENVIRONMENTAL EFFECTS ON HALO SHAPES \& DYNAMICS}
\label{environment}

\subsection{Halo Shape - Mass correlation}
In the $\Lambda$CDM cosmology, the dependence
of shapes on DM halo mass has been well established in many recent
studies with more massive haloes being less spherical, ie., 
having a lower axis ratio, $c/a$ 
(eg. Bullock 2002; Jing \& Suto 2002; Kasun \& Evrard 2005; 
Gottl\"ober \& Turchaninov 2006; 
Allgood et al. 2006; Paz et al. 2006; Macci\'o et al. 2006; Bett et al. 2006).
This can be explained considering that in the hierarchical clustering 
of CDM haloes, smaller mass haloes form earlier on average than massive ones, 
and thus they have more time to evolve, virialize and become more spherical.
We should also note that including baryonic physics has a significant
effect on the shapes of haloes (eg., Kazantzidis et al. 2004).
In this section we investigate whether this trend
changes when considering groups in different environments. Such a
difference has been noted by Avila-Reese et al. (2005) between galaxy size 
haloes found in clusters and in voids, and by
Hahn et al. (2006) between haloes found in
clusters and in filaments but only for small halo masses ($M\le
2\times 10^{12} M_\odot$). We do not probe this mass range and
therefore our analysis concentrates only on larger mass haloes, 
typical of groups and poor clusters of galaxies.

\begin{figure}
\epsfxsize=0.5\textwidth
\hspace*{-0.5cm} 
\centerline{\epsffile{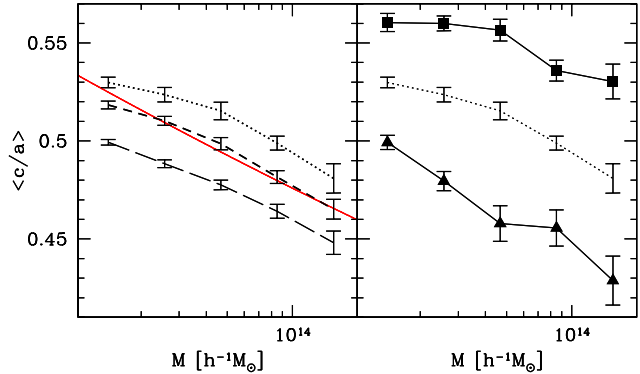}}
\caption{
{\em Left panel:} Mean axial ratios $\langle c/a \rangle$ as a function of halo
mass and environment. Dotted, short dashed and long dashed
lines correspond to halo samples of different cluster-group distances 
(H$_{0-7}$, H$_{10-17}$ and H$_{30-50}$, respectively).
The standard deviation of the axial ratio distribution for the
different halo subsamples and halo masses is between 0.11
and 0.12.
Error bars were calculated using the bootstrap re-sampling technique. 
The solid line denotes Allgood et al. (2005) fit. 
{\em Right panel:} Mean axial ratio $\langle c/a \rangle$, as a
function of halo mass, only for the H$_{0-7}$ haloes (dotted line). 
Squares and triangles correspond 
to H$_{0-7}$ haloes with $\Delta$-deviation values lower and higher than 
$\langle \Delta_{\rm bin}\rangle$, respectively. 
}
\label{allgood}
\end{figure}

Figure \ref{allgood} (left panel) presents our results for 
the three $r_{\rm cluster}$ subsamples (H$_{0-7}$, H$_{10-17}$ 
and H$_{30-50}$), with error bars computed using the bootstrap 
re-sampling technique. 
In all cases the trend we find is in accordance with the well known 
mass-shape relation, with $\langle c/a\rangle$ increasing with
decreasing halo mass.
However, it is shifted toward more spherical axial ratios 
when considering haloes nearer to massive clusters. 
This can be explained by the fact that haloes in 
high-density environments are formed earlier than haloes, of the same
mass range, in low-density environments (eg. Sheth \& Tormen 2004), 
giving the former more time to evolve, relax and hence become more spherical
(Avila-Reese et al. 2005).
This is in agreement 
with haloes at higher redshifts being more elongated than present day 
equal mass haloes (eg. Allgood et al. 2005). 

\begin{figure}
\epsfxsize=0.5\textwidth
\hspace*{-0.5cm} 
\centerline{\epsffile{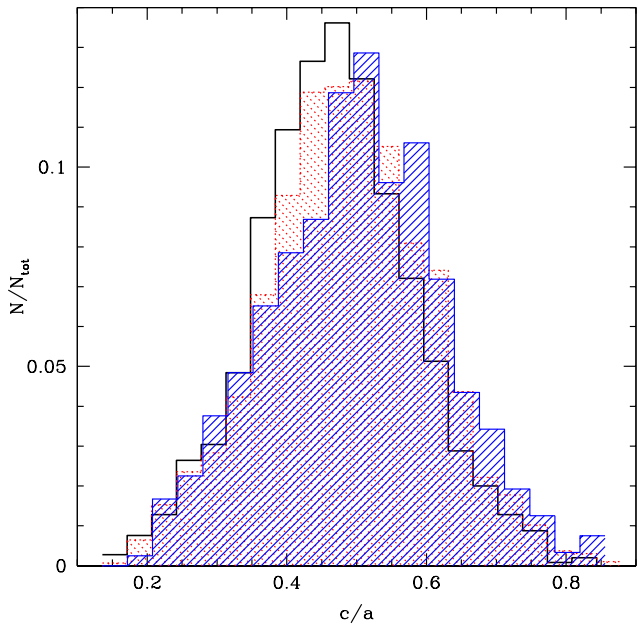}}
\caption{
Axial ratio ($c/a$) distributions for haloes in the mass range 
$3.8\times 10^{13} - 10^{14}\masa$.
Shaded, dotted and empty histograms correspond to the H$_{0-7}$, H$_{10-17}$ and
H$_{30-50}$ halo samples, whose means are 0.50, 0.48 and 0.47, respectively.
The $1\sigma$ dispersion is in all cases $\sim 0.1$ 
}
\label{histforma}
\end{figure}

Another representation of our results is shown in
Figure \ref{histforma} where we plot the $c/a$ frequency distribution
of well resolved haloes (ie., those with $3.8\times 10^{13}<
M< 10^{14}\masa$). The means of these distributions are
0.50, 0.48 and 0.47 for the H$_{0-7}$(dashed histogram), H$_{10-17}$ (dotted
histogram) and H$_{30-50}$ (empty histogram) samples, respectively,
while a Kolmogorov-Smirnov two-sample test shows them to be different at a very
high significant level.
%The $1\sigma$ dispersion is in all cases $\sim 0.1$ as in Allgood et al. 2005.

We now consider only the H$_{0-7}$ haloes and compute the shape-mass 
relation but separating haloes with high and low $\Delta$-deviation 
(substructure). Results are plotted in the right panel of 
Figure \ref{allgood}, where squares correspond 
to haloes with $\Delta<\langle \Delta_{\rm bin}\rangle$ 
and triangles to haloes with $\Delta>\langle \Delta_{\rm bin}\rangle$.
The shape-mass relation is basically maintained in both subsamples
although: 
(a) it is much shallower for haloes in the low $\Delta$-deviation subsample, and 
(b) it is shifted toward lower $\langle c/a \rangle$ for haloes in the 
high $\Delta$-deviation subsample. 
This result is indeed expected if to consider that more virialized
systems tend to be more spherical.
Those system with high level of substructure, which are dynamically younger 
systems, have interrupted their virialization process due to some
recent merger event and therefore have a more elongated shape than
systems with no sign of
substructure. This latter
behavior is present in the whole range of considered masses and also
in the H$_{10-17}$ and H$_{30-50}$ subsamples.

In order to discard the possibility that discreteness effects could impose
the $\langle c/a\rangle$-Mass
correlation, we recomputed the halo shapes but using 
realizations of only 130 randomly
selected particles per halo, so as to resemble the resolution of
the less massive haloes (see also Paz et al. 2006 for discreteness
related effects).
We indeed recover the same $\langle c/a \rangle$-Mass trend 
and thus we verify that it is not imposed by the variable resolution
with which the different mass haloes are resolved. However,
there is a shift toward less spherical
values when using the common halo resolution of 130 randomly selected 
particles per halo.
The apparent curvature, toward lower $c/a$ values and at the low-mass end 
of the $\langle c/a \rangle$-Mass relation, apparently disappears when
using the common halo resolution and thus it should probably be
attributed to the variable halo resolution.

\subsection{Halo Velocity Dispersion-Mass correlation}
\begin{figure}
\epsfxsize=0.5\textwidth
\hspace*{-0.5cm} \centerline{\epsffile{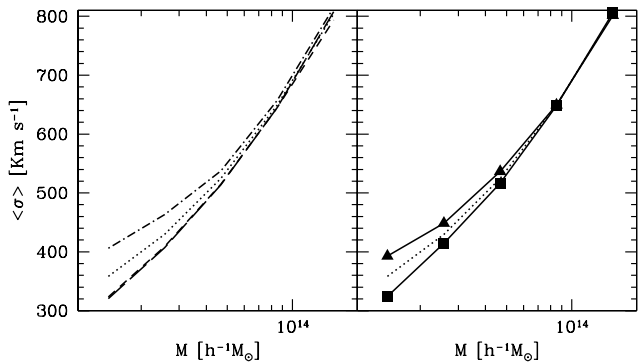}}
\caption{ Mean velocity dispersion $\langle\sigma\rangle$ as a
  function of halo mass. 
{\em Left panel:} Results for the usual three
$r_{\rm cluster}$ subsamples (H$_{0-7}$, H$_{10-17}$, H$_{30-50}$), 
with line types already
defined in Figure \ref{allgood}, while the 
dot dashed line corresponds to haloes with $r_{\rm cluster}<4 \; h^{-1}$
  Mpc (the H$_{10-17}$ and H$_{30-50}$ results are identical). 
Bootstrap errors are small, typically $\sim 10-20$ km/sec.
{\em Right panel:} Results only for the H$_{0-7}$ halo subsample
Squares correspond to haloes with $\Delta<\langle \Delta_{\rm bin}\rangle$ 
and triangles to haloes with $\Delta>\langle \Delta_{\rm bin}\rangle$ .
}
\label{sig_dist}
\end{figure}

In Figure \ref{sig_dist} we present the halo velocity dispersion-mass 
correlation. Velocity dispersions were computed using formula \ref{form-sig}. 
Such a correlation is expected from to the virial theorem. In order to
investigate the possible influence of the environment on this relation
and hence on the reliability of using the virial theorem to estimate
halo masses, we present in the left panel of Fig.\ref{sig_dist} results
for the three halo $r_{\rm cluster}$ (H$_{0-7}$, H$_{10-17}$ and H$_{30-50}$) 
subsamples and for an extra subsample with $r_{\rm cluster}<4 \mpc$ 
(dot-dashed line).

As expected from the virial relation we find that larger mass haloes have 
higher velocity dispersions, for all the considered subsamples. However, 
there is a shift toward higher velocity dispersions of low-mass haloes 
found near clusters with respect to those found further away. 
This trend is stronger the nearer the low-mass halo is found
to the cluster. 
However, for halo-cluster distances $\gsim 10 \mpc$ there is no
effect whatsoever. 

To investigate whether the halo dynamical state relates to the
halo velocity dispersion-mass correlation, we divide the H$_{0-7}$ subsample,
as in the previous section, to those with and without substructure.
Results are shown in the right panel of Figure \ref{sig_dist},
where the dotted line stands for all the H$_{0-7}$ haloes,
squares and triangles for haloes with $\Delta<\langle \Delta_{\rm bin}\rangle$ 
and $\Delta>\langle \Delta_{\rm bin}\rangle$ , respectively.
There is indeed a dependence of the mass-velocity dispersion 
correlation on the amount of halo substructure but only for haloes with masses
$\lsim 5\times 10^{13}\;h^{-1} \; M_\odot$, which are found to have 
a larger mean velocity dispersion than haloes with no or low-levels of
substructure. Note that Evrard et al. (2007) find similar results for what
they call satellite haloes.
Moreover, we find that haloes with a low $\Delta$-deviation index
behave similarly as haloes in the  H$_{10-17}$ and H$_{30-50}$ 
samples (seen in the left panel of Figure \ref{sig_dist}), which show no 
dependence of the mean velocity dispersion on the presence or not of
halo substructure.
Note, however, that a slight shift toward higher velocity dispersion
values is present also in high mass haloes but only if choosing those
haloes with extremely high $\Delta$-deviation index.

\subsection{Partial Conclusions}
These results, concerning the dependence of halo shapes and velocity
dispersion on the halo dynamical state,
give new insights in our understanding of halo formation and evolution.

Although the general expectation is that low-mass haloes in high
density environments formed earlier and thus should be relatively more
virialized with respect to similar mass haloes in 
low-density regions, we have found 
a relatively high fraction of dynamically young and active haloes near
massive clusters ($\gsim 45\%$).
These haloes  have in general a 
higher velocity dispersion (more evident at the low mass end) and a lower 
$\langle c/a\rangle$ ratio with respect to similar mass virialized haloes.
The high level of substructure  
of these haloes is probably because they are continuing to grow via mergers
in the anisotropic outskirts of 
massive haloes (eg. West 1994; Maulbetsch et al. 2006), 
although their dynamical state could also be affected from 
the strong tidal field imposed by their local high density
surrounding, while both cases imply a lower halo sphericity, as observed.
The mergers as the most possible cause for the increase 
of the halo velocity dispersion is in agreement with Faltenbacher et 
al. (2006) who find for an equal 
mass merging event (progenitors with masses $\sim 1\times10^{14}\masa$) 
an oscillatory behaviour of the velocity dispersion (among other properties).
After the relaxation of the new system, the velocity dispersion is slightly
larger, but it changes substantially during the event.

Now, higher mass haloes with a high $\Delta$-deviation (substructure) 
in the vicinity of massive clusters also appear to be of
lower sphericity although, and contrary to the low-mass halo case, their 
velocity dispersion does not show any significant 
deviation from that of the more virialized high-mass haloes. 
This could be explained if typically the merger events, which alter
the higher mass halo shape, are due to relatively lower mass haloes which
although affect the overall shape, they affect less the dynamical
structure of the high mass halo, which is dominated by the main
gravitational potential of the high-mass halo itself. It could also imply
a faster ``re-accommodation" of the velocity field with respect to the
density field, in the relatively deep principal halo potential well.

\section{ENVIRONMENTAL EFFECTS ON HALO ALIGNMENTS}
\label{alin}
\subsection{Internal Halo Alignments}
\label{internos}
In section \ref{introduccion}
we mentioned that there is a significant signal of alignment between
the principal axes of the shape and the velocity anisotropy tensors, 
indicating that most of the haloes 
have their shapes supported by the velocities of their member
particles (eg. Tormen 1997; Kasun \& Evrard 2005).

Here we investigate whether there is any environmental dependence 
on such an alignment effect.
To this end we compute the mean absolute cosine of the angle between
the major axes of the mentioned tensors, 
$\langle |{\bf \hat{a}} \cdot {\bf \hat{a}}_{\rm vel}|
\rangle$, as a function of the halo mass for the three halo subsamples
(H$_{0-7}$, H$_{10-17}$ and H$_{30-50}$).

\begin{figure}
\epsfxsize=0.5\textwidth
\centerline{\epsffile{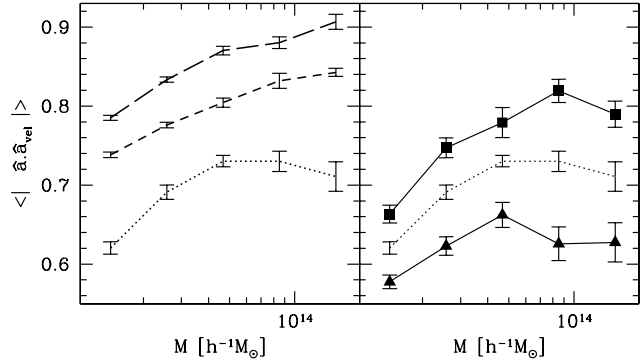}}
\caption{ 
{\em Left panel}: the direction cosine of the major axes of the shape and 
velocity ellipsoids 
$\langle |{\bf \hat{a}} \cdot {\bf \hat{a}}_{\rm vel} |\rangle$ as a 
function of halo mass for the H$_{0-7}$, H$_{10-17}$ and H$_{30-50}$ halo subsamples. 
Line styles are as in Figure \ref{allgood}.
Error bars were calculated using the bootstrap re-sampling technique.
{\em Right panel}: the corresponding 
$\langle |{\bf \hat{a}} \cdot {\bf \hat{a}}_{\rm vel}|\rangle$-mass 
correlation only for the H$_{0-7}$ sample, split in those haloes with high
level of substructure, $\Delta>\langle \Delta_{\rm bin}\rangle$  (triangles), 
and those without substructure, $\Delta<\langle \Delta_{\rm bin}\rangle$ 
(squares).}
\label{int_alig1}
\end{figure}

In the left panel of Figure \ref{int_alig1} we show the 
$\langle |{\bf \hat{a}} \cdot {\bf \hat{a}}_{\rm vel}|\rangle$-mass 
correlation for the H$_{0-7}$ (dotted line), H$_{10-17}$ (dashed line) 
and H$_{30-50}$ (long dashed line) samples. 
In all cases there is a good signal of alignment between the
shape and velocity ellipsoid principal axes, specially for haloes at
large distances from massive clusters
(H$_{10-17}$ and H$_{30-50}$ subsamples), 
while within each subsample the alignment is stronger
for the higher mass haloes. We have verified that this is not due to
the variable resolution with which the different mass haloes are
sampled.

Haloes at large distances from  massive clusters have their velocity and shape
better correlated probably because they are less tidally disrupted
than in the high density environment of the
cluster, either by the cluster itself or/and by the local overabundance of lower
mass haloes found in such environment. Even more so in the high-mass
halo end probably because interactions and merging with lower mass haloes
can disturb minimally the phase-space of these higher mass haloes. 
However, near
the massive cluster (H$_{0-7}$ subsample), the stronger halo-cluster 
gravitational 
interactions affect significantly the halo phase-space and for this reason we
observe a general decrease of the value of 
$\langle |{\bf \hat{a}} \cdot {\bf \hat{a}}_{\rm vel}|\rangle$ for all
halo masses.

This interpretation could be supported if those haloes with a high level
of substructure showed even less aligned orientations. Indeed this is the
case, as can be seen in the right panel
of Figure \ref{int_alig1}, where we present for the H$_{0-7}$ subsample the
$\langle |{\bf \hat{a}} \cdot {\bf \hat{a}}_{\rm vel}|\rangle$-mass 
correlation but split between haloes with high level of substructure, 
$\Delta>\langle \Delta_{\rm bin}\rangle$ (triangles),
and haloes with no substructure, $\Delta<\langle \Delta_{\rm bin}\rangle$  (squares).
The former haloes show a weaker alignment, as anticipated,
suggesting that strong interactions and mergers introduce scatter 
in the phase space of these systems.

The segregation between haloes with and without substructure,
seen in the right panel of Figure \ref{int_alig1}, 
is also present in the more distant halo samples (H$_{10-17}$ and
H$_{30-50}$). The same interpretation, given before, of the difference 
between equal mass haloes with and without substructure, holds 
for these samples as well. 
Furthermore, the slightly better alignment seen for the more massive haloes 
is due to their deeper potential wells, which inevitably 
creates a better alignment.

\begin{figure}
\epsfxsize=0.5\textwidth
%\hspace*{-0.5cm} 
\centerline{\epsffile{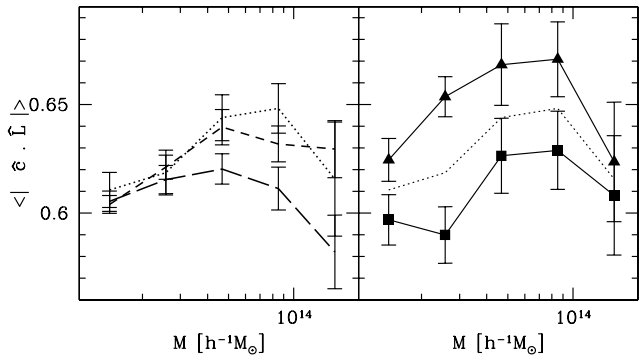}}
\caption{ 
{\em Left panel}: the direction cosine between the angular momentum
and the minor axis $\langle |{\bf \hat{c}} \cdot {\bf
  \hat{L}}|\rangle$ as a function 
of halo mass for the H$_{0-7}$, H$_{10-17}$ and H$_{30-50}$ halo subsamples. 
{\em Right panel}: the corresponding 
$\langle |{\bf \hat{c}} \cdot {\bf \hat{L}}|\rangle$-mass 
correlation only for the 
H$_{0-7}$ sample, split in those haloes with high level of substructure, 
$\Delta>\langle \Delta_{\rm bin}\rangle$  (triangles), 
and those without substructure, $\Delta<\langle \Delta_{\rm bin}\rangle$ 
(squares).}
\label{int_alig2}
\end{figure}

Another internal alignment effect that we address is that between
the directions of the angular momentum vector and the minor axis of 
the mass distribution, $|{\bf \hat{c}} \cdot{\bf \hat{L}}|$.
It has been found that the angular momentum is most often aligned with the 
minor axis and perpendicular to the major axis (eg. Dubinski 1992; 
Bailin \& Steinmetz 2005; Bett et al. 2006). An environmental dependence has also
been found by Avila-Reese et al. (2005) for galaxy-sized haloes, with
a higher alignment signal in underdense regions (see also Hahn et
al. 2006 for angular momentum orientations with large-scale structures).

Our results are shown in Figure \ref{int_alig2}, where we also find
such an alignment signal, although it is obvious
that due to noise we are unable to detect different trends 
in the three halo subsamples. Note also that the amplitude of our
alignment signal is significantly less than that found by Bailin \& Steinmetz
(2005), most probably because these authors define shapes using the reduced 
moment of inertia tensor, which weights strongly the inner parts of
haloes, as well as because they choose to analyse only haloes of
which both their small axis and angular momentum orientation have small
uncertainties. Furthermore, one should keep in mind that, due to
resolution effects (see section 3.2), the intrinsic
uncertainty of this alignment measure is quite large for low-mass haloes.

Returning to our results we do find a systematic, although weak, trend of 
a better alignment for the H$_{0-7}$ subsample (left panel of Figure
\ref{int_alig2}), which is in the opposite direction than the results
of Avila-Reese et al. (2005) based on galaxy-sized haloes. 
However, our haloes are much larger
(groups and poor cluster size) and this could well be the reason of
the apparent discrepancy. Furthermore, we are in general agreement with the
recent results of Arag\'on-Calvo et al. (2006).

Moreover, haloes with high level of substructure 
seem to have ${\bf \hat{L}}$ and ${\bf \hat{c}}$ better correlated
(right panel Figure \ref{int_alig2}), a fact which is true for all halo
subsamples (H$_{0-7}$, H$_{10-17}$ and H$_{30-50}$). This should be
partly attributed to the fact that haloes with substructure are more elongated
than relaxed haloes and thus they have both their angular
momentum and minor axis vectors better defined.
Furthermore, this result also implies that mergers probably
affect significantly the angular momentum
of haloes (eg. Vitvitska et al. 2002), which gain part of their angular momentum
from mergers preferentially occurring along the plane defined by the major and 
median axes.

\subsection{External Alignments}
It has been shown that the orientation of the halo major axis
is strongly correlated with the direction from which
the last major merger event occurred \cite{van}.
Therefore, it should be expected to find a correlation between halo major
axis orientation and the direction defined by the halo-cluster distance,
which in turn should
indicate the orientation of the filament. Such alignment effects,
among relatively massive haloes, have been found to be particularly strong 
and extending up to
$\gsim 100 \; h^{-1}$ Mpc (eg., Faltenbacher et al. 2002; Kasun \& Evrard
2005; Hopkins, Bachall \& Bode 2005).

\begin{figure}
\epsfxsize=0.5\textwidth
%\hspace*{-0.5cm} 
\centerline{\epsffile{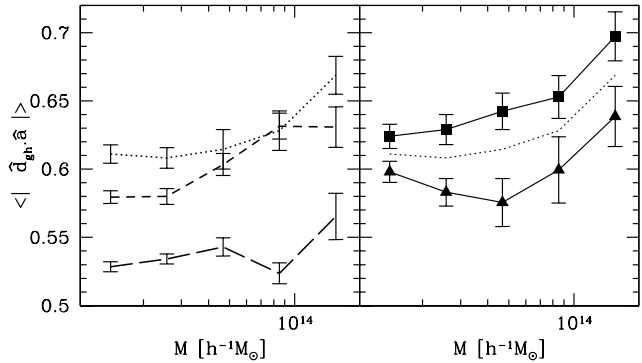}}
\caption{
{\em Left panel:} Correlation between halo-cluster direction and 
halo major axis orientation 
(${\rm \langle |{\bf \hat{d}}_{gh} \cdot {\bf \hat{a}}|\rangle}$) 
as a function of halo mass.
Long dashed, short dashed and dotted lines correspond to 
different bins of cluster-group distances as in Figure \ref{allgood}.
Error bars were calculated using the bootstrap re-sampling technique.
{\em Right panel:} Similar but only for the H$_{0-7}$ sample, split into 
 haloes with and without significant substructure, $\Delta>\langle
 \Delta_{\rm bin}\rangle$  (triangles) and  $\Delta<\langle
 \Delta_{\rm bin}\rangle$ (squares), respectively.}
\label{ext_alig1}
\end{figure}

\begin{figure}
\epsfxsize=0.5\textwidth
%\hspace*{-0.5cm} 
\centerline{\epsffile{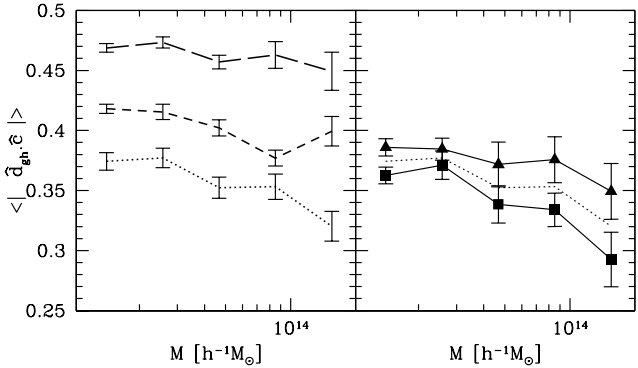}}
\caption{
{\em Left panel:} Correlation between cluster-group direction 
and halo minor axis orientation 
(${\rm \langle |{\bf \hat{d}}_{gh} \cdot {\bf \hat{c}}|\rangle}$) 
as a function of halo mass.
{\em Right panel:} Similar but only for the H$_{0-7}$ sample, split into 
 haloes with and without significant substructure.}
\label{ext_alig2}
\end{figure}

Here we also wish to calculate the alignment between the direction 
of each halo to their
nearest cluster (${\rm {\bf \hat{d}}_{gh}}$) 
and the halo major or minor axis orientation
(${\rm \langle |{\bf \hat{d}}_{gh} \cdot {\bf \hat{a}}|\rangle}$ 
and ${\rm \langle |{\bf \hat{d}}_{gh} \cdot {\bf \hat{c}}|\rangle}$,
respectively) for all three (H$_{0-7}$, H$_{10-17}$ and H$_{30-50}$) 
halo subsamples.
The left panel of Figure \ref{ext_alig1} 
shows the case for the
major axis alignment while the left panel of Figure \ref{ext_alig2} shows the
corresponding minor axis alignment case. As expected, the former 
alignment effect is strong, more so for haloes found near their clusters
and for the high mass haloes.
As in Bailin \& Steinmetz (2005) we find that the halo minor axes are 
in general anti-aligned (perpendicular) to the filament direction, 
again more so
for haloes found near their clusters and for high mass haloes. 
Now dividing our H$_{0-7}$ subsample into those haloes with and without
substructure (right panels of Figures \ref{ext_alig1} and \ref{ext_alig2}), 
we find very interesting results:

\begin{itemize}
\item relatively virialized haloes, having no significant substructure, show
a strong tendency for major axis alignment (and minor axis anti-alignment) 
with the direction to their nearest massive cluster implying that they
retain strong memory of the initial anisotropic distribution from
which they accreted matter (eg. van Haarlem \& van de Weygaert 1993), 
\item haloes with high level of substructure show a similar, although
  relatively weaker, alignment effect, which appears to be in 
disagreement with what one would naively expect given that 
mergers happen preferentially along the
filaments. However, once a merger has happened, non-linear
gravitational effects take place and until the merged structure
relaxes, it may well appear less aligned with the filament
orientation. Specially, low mass haloes, being also small in size,
interacting with other neighboring haloes
in the high density surroundings of a massive cluster, could be
relatively more
affected by local gravitational effects which may not necessarily
reflect the large-scale anisotropic distribution of matter in the filament.
\end{itemize}

An interesting complication in the above interpretation is that
the halo angular momentum and minor axis is better aligned 
in high $\Delta$-deviation haloes (see section
\ref{internos}) and 
if directional mergers are responsible for such an alignment, being more 
frequent in the filament direction, then one might have expected their major axes 
to be more aligned  with the halo-cluster direction than for
virialized haloes (small $\Delta$-deviations).
However, what the angular momentum-minor axis halo alignment actually
implies, in the above picture, is
that the direction of the merger is in the plane defined by the
major and medium axes and not necessarily along the major axis, which
is not in contradiction with the above.

\section{Discussion \& Conclusions}
\label{conclusiones}
We have used haloes identified with a FOF algorithm in a dark matter only 
$\Lambda$CDM simulation to study the dependence of the shape,
dynamical state and various
alignments of group size haloes on the halo environment. The smallest
haloes analysed have at least 130 particles ($M\gsim 10^{13}\masa$). 
In order to investigate if there is some influence of the environment
on the properties of haloes, we split the group sample in three
subsamples according to their distances to the nearest massive {\em cluster} 
halo.
We have also investigated the Dressler \& Shectman (1996) algorithm,
used to determine whether a halo has a high level of substructure or not,
considered as an indication of their dynamical state, and
we devised a substructure characterization of the haloes 
which is free of halo resolution effects. We then
divide the haloes to those with and without a high level of substructure.

Our results can be summarized in the following:
\begin{enumerate}
\item The well known relation between halo shape and halo mass
has also an environmental dependence, albeit weak. 
Haloes found at small distances from their
massive cluster show a systematic shift toward larger axial ratios (ie.,
they are more spherical) with respect to equal mass haloes at larger 
distances, a fact which is true for the whole range of halo masses 
studied. This result appear to be in disagreement with Hahn et al. (2006).

\item The velocity dispersion of equal mass haloes shows a dependence
  on the environment. Haloes with substructure, near massive clusters, 
have a larger velocity dispersion with respect to equal mass haloes 
with no or low substructure index. This is probably due to the higher
  halo merging rate in high-density environments.
On the contrary, haloes found further away do not exhibit this same behaviour:
the velocity dispersion-mass trend is the same independent of the
  presence or not of substructure. 
The velocity dispersion of high mass haloes does not seem to be 
affected by the environment, nor that of any halo at a distance 
larger than $\sim 10 \mpc$ from its nearest massive cluster. 

\item The influence of environment is also reflected in the internal alignment 
of the velocity and density ellipsoid principal axes.
Such an alignment is stronger for higher mass groups,
probably due to the better definition of their shape given that
these groups are more elongated than lower mass ones,
while it is weaker near massive clusters, where the influence 
of the cluster and the high-density halo
neighborhood is stronger. It is even weaker for haloes with a high level
of substructure, which  reflects the fact that during a merger the
halo phase-space is significantly perturbed.

\item Angular momentum and minor axes of haloes are roughly aligned, 
even more so for haloes with substructure. 
This relation does not seem to depend strongly on
environment. However, one should keep in mind that the uncertainty
of this measure is quite large, due to resolution effects.

\item On larger scales we detect alignments between the orientation of
  a halo and the direction to its nearest massive cluster, which probably
  reflects the orientation of the filament within which they are embedded.
The halo minor/major axes  appear perpendicular/parallel to the
  filament, while the signal for both alignments is stronger for
  haloes near massive clusters and for haloes with no substructure.

\item Overall we have found that the halo properties studied in this
  work as a function of the distance to their nearest cluster, 
show a strong dependence on the amount of halo substructure.
Since significant halo substructure is related to on-going 
or a recent merger, we could infer
that the influence of the close neighborhood of a halo, in the
vicinity of massive clusters, is not less
important than the influence of the cluster itself.
\end{enumerate}

There are at least two mechanisms involved in the evolution
of the shape, alignment and velocity dispersion of haloes, namely
the formation time (eg. Gottl\"ober, Klypin \& Kratsov 2001; Sheth \& Tormen
2004) and the influence of the immediate environment. In
this paper we were concerned with the latter aspect of the problem.
We can summarize the interpretation of our results as follows.

On the one hand, haloes forming in high density regions collapse earlier and
they would on average have had more time to evolve and thus
sphericalize, more so with respect to equal mass haloes forming in low-density
regions. However, in the high-density environments an opposing factor 
is the overabundance of haloes which induce mergers and intra-halo
interactions, which then disturb the virialized nature of these older
haloes.
Higher mass haloes evolve hierarchically by the
accretion of lower-mass haloes and thus are more elongated with respect
to lower-mass haloes, which collapse and form earlier according to
CDM models.

The rising of the velocity dispersion of haloes (having significant
substructure) near massive clusters, with
respect to equal mass haloes with insignificant substructure 
or with those found in lower density regions, 
could be attributed to 
the higher halo merging rate present in the high-density environment 
of massive clusters.
The fact that the fraction of haloes with significant substructure
is higher in high-density regions (see end of section
\ref{formaysubs}), indeed reflects the more frequent halo mergers and 
interactions, which introduces also a bulk-flow (infall) velocity 
component in the halo velocity dispersion measure. 

The merging processes in high-density environments occur along the
anisotropic distribution of matter, which defines the large-scale
filaments orientation. This is reflected in the alignments
of the angular momenta of haloes, which are strongly influenced by the
merging process, with the minor axis of the halo and the alignment of
the halo major axis with the orientation of the filament, defined by
the direction between the halo and its nearest massive cluster.

\section*{Acknowledgments}
C.J. Ragone-Figueroa is supported by a CONICET fellowship, Argentina.
This work has been partially supported by the European Commission's 
ALFA-II programme through its funding of the Latin-american European Network 
for Astrophysics and Cosmology (LENAC), 
the Consejo de Investigaciones Cient\'{\i}ficas y T\'ecnicas de la Rep\'ublica 
Argentina (CONICET), the Secretar\'{\i}a de Ciencia y T\'ecnica de la 
Universidad Nacional de C\'ordoba (SeCyT), Agencia Nacional de Promoci\'on 
Cient\'{\i}fica de la Rep\'ublica Argentina and Agencia C\'ordoba Ciencia.

\end{document}